\documentclass[twocolumn,nofootinbib]{revtex4-2}
\usepackage{graphicx}
\usepackage{color}
\usepackage{enumerate}
\usepackage{amssymb}
\usepackage{amsmath}
\usepackage{amsfonts}
\usepackage{bbm}
\usepackage{dsfont}

\begin{document}

\title{Visualizing quantum coherence and decoherence in nuclear reactions}

\author{K. Hagino}
\affiliation{ 
Department of Physics, Kyoto University, Kyoto 606-8502,  Japan} 

\author{T. Yoda}
\affiliation{ 
Department of Physics, Kyoto University, Kyoto 606-8502,  Japan} 

\begin{abstract}
Differential cross sections of nuclear reactions often exhibit characteristic oscillations in the angular distribution 
originated from an interference of two indistinguishable processes. 
Here we propose a novel method to visualize origins of such oscillations. 
This is achieved by taking Fourier transform of scattering amplitudes, following the idea 
in wave optics. We apply this method to elastic scattering of $^{16}$O+$^{16}$O and 
$^{18}$O+$^{18}$O at energies above the Coulomb barrier. 
The former system shows strong oscillations in the angular distribution due to 
the nearside-farside interferences, while the oscillations are largely suppressed 
in the latter system due to a stronger absorption.  
We show that the image of the former and the latter systems corresponds 
to a double-slit 
and a single-slit problems in quantum mechanics, respectively. 
\end{abstract}

\maketitle

In quantum mechanics, when two or more indistinguishable processes are involved, 
the probability is computed by taking the absolute square of the total amplitude, which 
is given as a sum of the amplitude of each process. 
This leads to the interference of each process due to the cross terms. 
This is referred to as quantum coherence, and this is one of the most fundamental 
features of quantum 
mechanics. In addition to the famous double-slit problem, 
a textbook example for this is scattering of two identical particles, for which 
a detector cannot distinguish scattering at angle $\theta$ from scattering at angle 
$\pi-\theta$. In this case,  
the differential cross sections are given by 
$d\sigma/d\Omega=|f(\theta)\pm f(\pi-\theta)|^2$, where 
$f(\theta)$ and $f(\pi-\theta)$ are scattering 
amplitudes for the angles $\theta$ and $\pi-\theta$, respectively, 
and the sign of the superposition depends on 
the statics of the particles. Due to the interference between 
$f(\theta)$ and $f(\pi-\theta)$, the differential cross sections exhibit characteristic oscillations  
as a function of the scattering angle $\theta$. 
Such oscillations have been actually observed e.g., in elastic scattering of 
$^{16}$O+$^{16}$O at energies below the Coulomb barrier\cite{bromley1961}. At such energies, 
the nuclear 
effect can be neglected, and the experimental data can be well accounted for 
by taking a superposition of the Rutherford 
scattering amplitudes at $\theta$ and $\pi-\theta$. 

Besides the interference due to the exchange of two identical particles, there are 
many other interference phenomena known in low-energy nuclear reactions. 
These include, the Coulomb-nuclear interference \cite{brink1985}, the nearside-farside 
interference \cite{fuller1975,rowley1976,hussein1984}, 
and the barrier-wave-internal-wave interference\cite{brink1977}. 
In particular, an analogy between the nearside-farside interference and the double-slit problem 
has been discussed in Ref. \cite{hussein1984}. 
Here, the nearside component corresponds to scattering at a positive scattering angle with a positive 
impact parameter while the farside component corresponds to scattering with a negative 
impact parameter. Due to a strong absorption inside a nucleus, scattering takes place only at the edge 
of a nucleus, which corresponds to scattering through two slits in a double slit problem. 

In this paper, we propose a novel way to visualize an origin of oscillations in the angular distribution 
of nuclear reactions. The idea of this method 
is to take Fourier transform of a scattering amplitude, similar to 
what is done in wave optics. A similar method has been applied in particle physics, in which 
the images of string scattering \cite{hashimoto2023} and that of 
black holes in the AdS/CFT correspondence \cite{hashimoto2019,hashimoto2020} 
have been discussed. 
In particular, it has been demonstrated that the image of string scattering corresponds to a double slit 
\cite{hashimoto2023}. 
Here we apply a similar method to elastic scattering of 
$^{16}$O+$^{16}$O and 
$^{18}$O+$^{18}$O at energies above the Coulomb barrier, 
and show how the quantum coherence in $^{16}$O+$^{16}$O is decohered in  
$^{18}$O+$^{18}$O by nuclear absorption. 

Following Ref. \cite{hashimoto2023}, we take an image of scattering 
using a lens located at the direction $(\theta_0,\varphi_0)$ from the scattering center. 
To this end, we take Fourier transform of scattering 
amplitude in a form of 
\begin{eqnarray}
\Phi(X,Y)&=&\frac{1}{S}\int^{\theta_0+\Delta\theta}_{\theta_0-\Delta\theta}
d\theta e^{ik(\theta-\theta_0)X} f(\theta)
\nonumber \\
&& \times\int^{\varphi_0+\Delta\varphi}_{\varphi_0-\Delta\varphi}
d\varphi\,
e^{ik(\varphi-\varphi_0)Y},
\label{eq:image}
\end{eqnarray}
where $(X,Y)$ is the coordinate on the virtual screen 
behind the lens and $k$ is 
the wave number, $k=\sqrt{2\mu E/\hbar^2}$, $\mu$ and $E$ being the reduced mass and the 
energy in the center of mass frame, respectively. 
See Supplemental Material for a derivation of this formula. 
We have assumed that the scattering amplitude $f$ is 
independent of the angle $\varphi$. In Eq. (\ref{eq:image}), $S$ is the angular area of the lens given by 
\begin{equation}
S=\int^{\theta_0+\Delta\theta}_{\theta_0-\Delta\theta}
d\theta\int^{\varphi_0+\Delta\varphi}_{\varphi_0-\Delta\varphi}
d\varphi\,
= 4(\Delta\theta)(\Delta\varphi). 
\end{equation}
The actual image is given by $I(X,Y)=|\Phi(X,Y)|^2$. 
In this paper, following Ref.\cite{hashimoto2023}, we take $\Delta\varphi=\Delta\theta/\sin\theta_0$, that corresponds 
to a square lens. 

Since the scattering amplitude does not depend on the angle $\varphi$, the integral for $\varphi$ 
is trivial in Eq. (\ref{eq:image}) and is given by 
\begin{equation}
\int^{\varphi_0+\Delta\varphi}_{\varphi_0-\Delta\varphi}
d\varphi\,
e^{ik(\varphi-\varphi_0)Y}
=2\Delta\varphi\,\frac{\sin(k Y\Delta\varphi)}{k Y\Delta\varphi}.
\label{eq:varphi}
\end{equation}
This function is peaked at $Y=0$ and has a width of $2\pi/(k \Delta \varphi)$\cite{hashimoto2023}. 
The resolution of the image in the $Y$ direction is thus determined by the quantity $k\Delta \varphi$. 
Notice that Eq. (\ref{eq:varphi}) is independent of $\varphi_0$. 
For a flat angular distribution with $f(\theta)$=const., the same argument holds for the position and the resolution of the peak
in the $X$ direction. 

Let us first apply Eq. (\ref{eq:image}) to 
Rutherford scattering, that is, scattering with a pure Coulomb potential, $V(r)=Z_1Z_2e^2/r$, 
where $Z_1$ and $Z_2$ are atomic numbers of two colliding nuclei, for which the scattering 
amplitude $f(\theta)$ is known analytically, see e.g. Ref. \cite{konishi2009}. 
Figure 1 shows the image of Rutherford scattering for $^{16}$O+$^{16}$O at $E_{\rm c.m.}=8.8$ MeV. 
Even though this is a system with identical bosons, the symmetrization of the 
wave function is not 
taken into account here. For the image, we take $\theta_0$=90 degrees with $\Delta\theta=\Delta\varphi$=30 degrees.
One can see that the image has a peak at $X=5.65$ fm. This is actually close to the classical 
impact parameter for Rutherford scattering of 
this system at $\theta_0$=90 degree, $b_{\rm cl}=5.24$ fm. 
As we derive in Supplemental Material, for Rutherford scattering the peak of the image 
indeed coincides with 
the classical impact parameter in the limit of $\Delta\theta\to 0$. 
We expect that this holds in general for heavy-ion reactions, for which the Coulomb interaction plays an 
important role in determining the reaction dynamics. 

\begin{figure}[t]
\includegraphics[width=7.5cm]{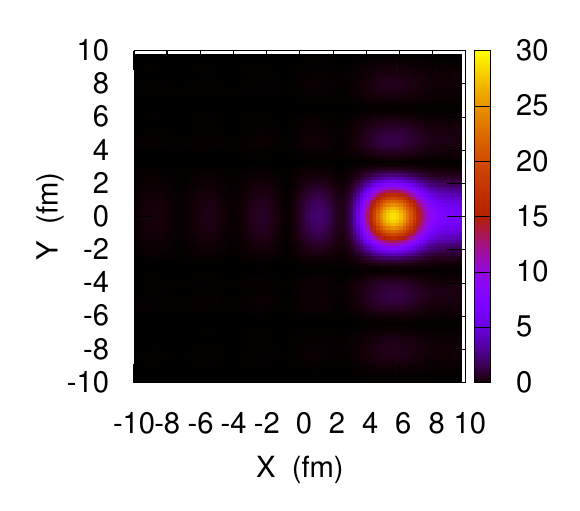}
\caption{
The image of Rutherford scattering 
for $^{16}$O+$^{16}$O at $E_{\rm c.m.}=8.8$ MeV with the 
unsymmetrized Coulomb scattering amplitude. 
The angles in Eq. (\ref{eq:image}) are set to be 
$\theta_0$=90 degrees and $\Delta\theta$=$\Delta\varphi$=30 degrees. 
}
\end{figure}

Notice that Eq. (\ref{eq:image}) for $\theta_0=\pi/2$ has a property that 
$\Phi(X,Y)$ with $f(\pi-\theta)$ is identical to  $\Phi(-X,Y)$ with $f(\theta)$. 
Therefore the image of $^{16}$O+$^{16}$O scattering with the symmetrized 
scattering amplitude $f(\theta)+f(\pi-\theta)$ 
has two symmetric peaks at $X_{\rm peak}$ and $-X_{\rm peak}$, just 
as in the double-slit problem discussed in Ref. \cite{hashimoto2023}. 
See the Supplemental Material for details. 

\begin{figure}[t]
\includegraphics[width=7.5cm]{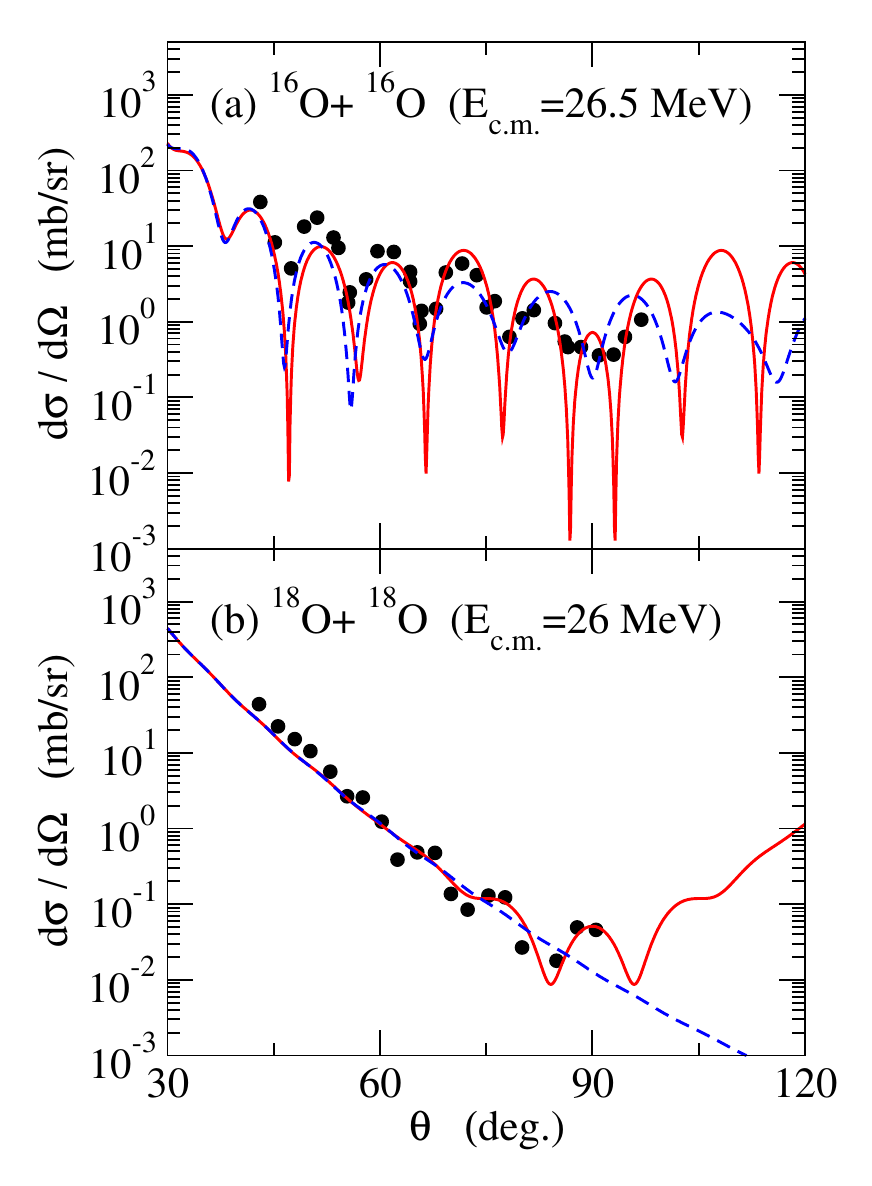}
\caption{
(the upper panel) The angular distribution of the $^{16}$O+$^{16}$O 
elastic scattering at $E_{\rm c.m.}=26.5$ MeV. The solid line shows 
a fit with a deep squared Woods-Saxon potential, while the dashed line shows 
the unsymmetrized cross sections obtained with the same potential. 
The experimental data are taken from Ref. \cite{16O16O}. (the lower panel) 
The same as the upper panel, but for the $^{18}$O+$^{18}$O 
elastic scattering at $E_{\rm c.m.}=26$ MeV. 
The surface imaginary potential is also added to the optical potential. 
The experimental data are taken from Ref. \cite{18O18O}. 
}
\end{figure}

Let us next discuss elastic scattering of $^{16}$O+$^{16}$O and 
$^{18}$O+$^{18}$O at energies above the Coulomb barrier, at which both the Coulomb and the 
nuclear interactions play a role. The upper panel of Fig. 2 shows the angular distribution 
of $^{16}$O+$^{16}$O elastic scattering at $E_{\rm c.m.}=26.5$ MeV \cite{16O16O}. 
With a standard global nuclear 
potential,  
the height of the Coulomb barrier for this system is estimated to be around 10.3 MeV, and thus this energy is about 2.6 times the barrier height. 
The experimental angular distributions for this system 
show a strong oscillatory pattern. 
We fit this with a deep squared Woods-Saxon potential for the nuclear part of internucleus potential \cite{kondo1989,ohkubo2002},
\begin{equation}
V_N(r)=-V_0\,g(R_R,a_R,r)^2-iW_0\,g(R_W,a_W,r)^2,
\label{eq:WS2}
\end{equation}
with 
\begin{equation}
g(R,a,r)=1/(1+\exp[(r-R)/a]).
\end{equation}
The solid line in the upper panel is obtained with the parameters 
$V_0$=421.28 MeV, 
$R_R=4.12$ fm, $a_R=1.52$ fm, $W_0$=157.1 MeV, $R_W=4.39$ fm, and 
$a_W=$ 0.151 fm, together with the radius of the uniform charge distribution 
of 5.54 fm. 
The observed oscillations are reasonably well accounted for with this parameter 
set. The lower panel of Fig. 2 shows the angular distribution for the 
$^{18}$O+$^{18}$O at a similar energy as the one shown in the upper panel for 
$^{16}$O+$^{16}$O \cite{18O18O}. For this system, the oscillatory pattern is much 
less pronounced (see also Ref. \cite{18O18O-2}), 
and the same squared Woods-Saxon potential 
as that for the 
$^{16}$O+$^{16}$O system 
does not fit well the 
experimental data. 
This is most likely due to the two extra neutrons outside the doubly magic 
$^{16}$O nucleus, with which the $^{18}$O nuclei are excited more easily 
than the 
$^{16}$O nuclei. A stronger absorption is necessary to fit 
the data\cite{charzewski1978,haas1981}, 
and for this purpose we introduce a surface imaginary potential, 
\begin{equation}
W_S(r)=-iW_s\,dg(R_s,a_s,r)/dr.
\end{equation}
The solid line in the lower panel is obtained with the parameters 
$W_s=94.01$ MeV, $R_s=5.61$ fm, and $a_s=0.734$ fm, together 
with the potential given by Eq. (\ref{eq:WS2}) with a scaling of $R_R$ and $R_W$ 
by a factor of 1.04 to account for the mass number dependence of the nuclear radii. 
One can see that this calculation well accounts for the data 
for the $^{18}$O+$^{18}$O system. 

\begin{figure}[t]
\includegraphics[width=7.5cm]{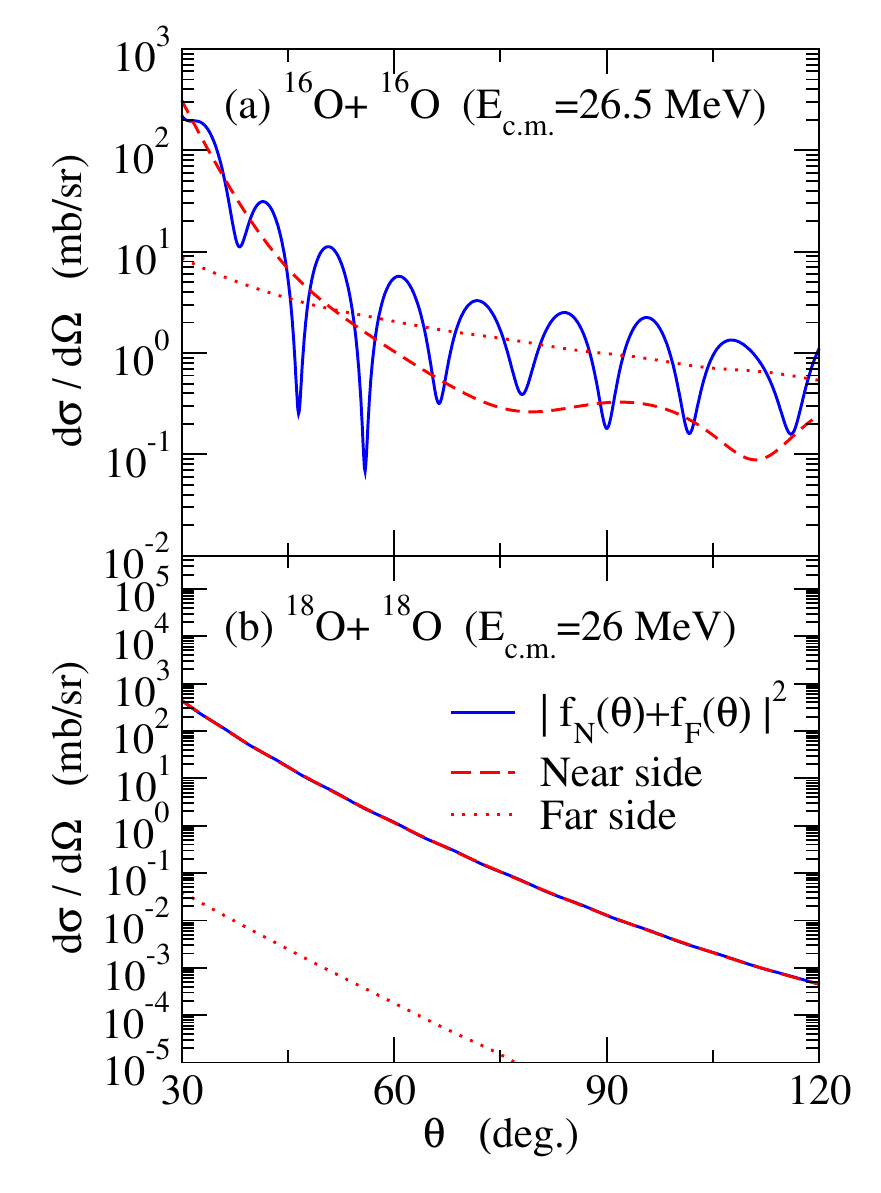}
\caption{
The unsymmtrized cross sections for elastic scattering of the 
$^{16}$O+$^{16}$O (the upper panel) and the 
$^{18}$O+$^{18}$O (the lower panel) systems. The solid lines 
show the total cross sections, while the dashed and the dotted lines 
denote their decompositons into the nearside and the farside components, 
respectively. 
}
\end{figure}

We notice that the unsymmetrized cross sections, obtained only with the scattering 
amplitude $f(\theta)$, show strong oscillations for the $^{16}$O+$^{16}$O 
system (see the dashed line in the upper panel). This indicates that the effect of 
symmetrization due to the identical bosons play a minor role at this energy 
in the oscillations 
for the $^{16}$O+$^{16}$O system, even though the small oscillations around $\theta=\pi/2$ 
for the $^{18}$O+$^{18}$O system is certainly due to the symmetrization of the 
wave function. 
To investigate the origin for the oscillations in the $^{16}$O+$^{16}$O, 
we decompose the scattering amplitude into the nearside and the farside 
components by using the Legendre functions of the second kind \cite{fuller1975}. 
The solid lines in Fig. 3 show the unsymmetrized cross sections for the 
$^{16}$O+$^{16}$O (the upper panel) and the 
$^{18}$O+$^{18}$O (the lower panel) systems, while the dashed and the dotted lines 
show their decompositions into the nearside and the farside components, respectively. 
The upper panel indicate that 
the nearside and the farside components cross each other at 
around $\theta=51$ degrees, 
and the strong oscillations are indeed 
caused by an interference between the nearside and the farside components. 
On the other hand, the farside component is largely suppressed in the 
$^{18}$O+$^{18}$O system due to the strong absorption, 
and the scattering amplitude is almost solely given by the nearside component. 
In this way, the quantum coherence observed in the 
$^{16}$O+$^{16}$O system is decohered in the $^{18}$O+$^{18}$O system due to the 
couplings to the internal degrees of freedom, that may be regarded as an internal 
environment. 

\begin{figure}[t]
\includegraphics[width=7.5cm]{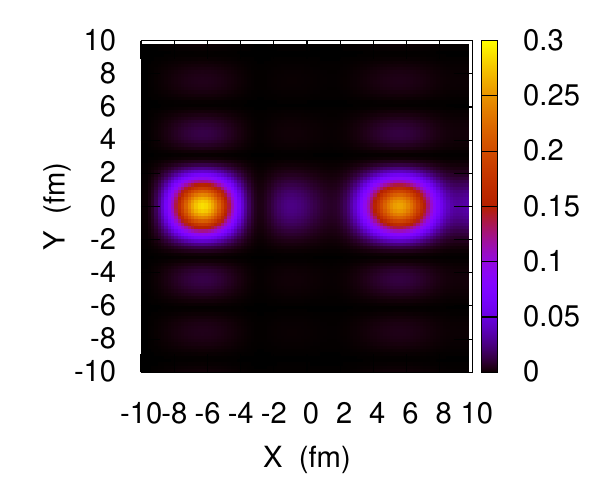}
\includegraphics[width=7.5cm]{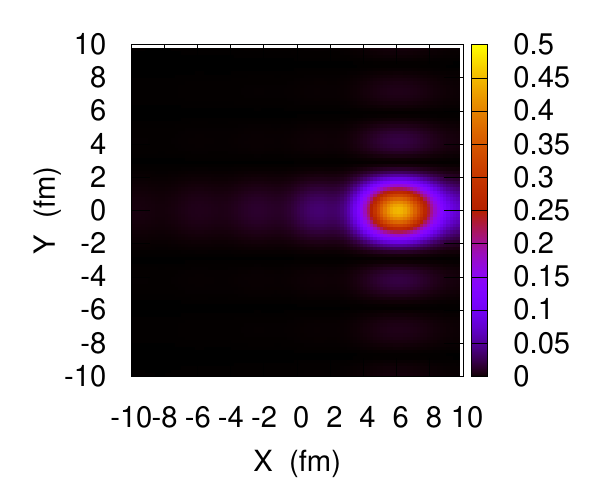}
\caption{
The images of the 
unsymmtrized cross sections for elastic scattering of the 
$^{16}$O+$^{16}$O (the upper panel) and the 
$^{18}$O+$^{18}$O (the lower panel) systems. 
The angles in Eq. (\ref{eq:image}) are set to be 
$\theta_0$=55 degrees and $\Delta\theta$=15 degrees. 
}
\end{figure}

The images of the unsymmetrized cross sections for the 
$^{16}$O+$^{16}$O and the $^{18}$O+$^{18}$O systems are shown in the upper and the 
lower panels of Fig. 4, respectively. 
These are obtained with $\theta_0$=55 degrees and $\Delta\theta$=15 degrees. Here we set $\theta_0$ close to the crossing 
point of the nearside and the farside components in the $^{16}$O+$^{16}$O 
so that both the components contribute with similar magnitudes. 
For the $^{16}$O+$^{16}$O system, the image has two distinct peaks. The analysis with 
the nearside and the farside amplitudes indicates that the peak at a positive 
value of $X$ corresponds to the nearside component while the peak at a negative $X$ 
corresponds to the farside component. 
As we have discussed with the Rutherford scattering, the peak of an image corresponds 
to the classical impact parameter of scattering. 
Fig. 4 therefore agrees with the physical picture of the nearside and the farside 
components, that is, the 
nearside and the farside components correspond to a positive and a negative 
impact parameters, respectively. 
This can also be seen in the $^{18}$O+$^{18}$O system, for which only the nearside 
component contributes significantly to the cross sections. 
The image for this system has a peak at a positive value of $X$, 
reflecting a positive impact parameter for the nearside component. 

In summary, we have proposed a novel way to image nuclear reactions.  
Based on an idea in wave optics, 
as had been 
advocated in the field of particle physics, 
the image can be obtained by performing 
Fourier transform of a scattering amplitude. 
For an angle-independent scattering amplitude, the image is peaked at the 
origin with the widths determined by parameters in the Fourier transform. 
For Rutherford scattering, the peak of the image is shifted to the position 
corresponding to the classical impact parameter of scattering. 
We have applied this method to elastic scattering of the $^{16}$O+$^{16}$O 
and $^{18}$O+$^{18}$O systems at energies about 2.6 times the Coulomb barrier. 
The image for the $^{16}$O+$^{16}$O system has been found to have 
two peaks, corresponding to the nearside and the farside components of 
the reaction process. The quantum interference between the two components is 
largely decohered in the $^{18}$O+$^{18}$O system due to the strong absorption 
originated from the two extra neutrons outside $^{16}$O. 
The image for this systems has been found to have a single peak, corresponding solely 
to the nearside component. 
Elastic scattering for 
the $^{16}$O+$^{16}$O 
and $^{18}$O+$^{18}$O systems at these energies therefore have close analogies 
to a double slit and a single slit problems in quantum mechanics, respectively. 

In this way, the imaging proposed in this paper provides an intuitive 
understanding of the origin and the underlying dynamics of quantum interference 
phenomena in nuclear reactions. 
Of course, a scattering amplitude is not an observable, unlike cross sections. 
However, one can make an attempt to fit data with an optical model, from which 
one can obtain a scattering amplitude to be used for imaging. 
There are a variety of interference phenomena in nuclear reactions. 
We leave applications of the imaging to these phenomena for interesting future works. 
An application to inelastic scattering \cite{rowley1984} will be another interesting direction 
for future works. 

\medskip

We thank Koji Hashimoto for useful discussions. 
This work was supported in part by JSPS KAKENHI
Grants No. JP19K03861 and No. JP23K03414.
The work of T.~Y.\ was supported in part by JSPS KAKENHI Grant No.~JP22H05115 and JP22KJ1896.

\bibliography{ref}
\bibliographystyle{apsrev4-2}

\end{document}


\setcounter{equation}{6}
\setcounter{figure}{4}

\section{supplemental material}

\subsection{Derivation of Eq. (1)}

In a scattering problem, one considers the asymptotic wave function in a form of 
\begin{equation}
\psi(\vec{r})\to e^{ikz}+f(\theta)\frac{e^{ikr}}{r}~~~~~(r\to\infty), \\
\end{equation}
where $k=\sqrt{2\mu E/\hbar^2}$ is the wave number with $E$ and $\mu$ being the energy in the center of mass frame and 
the reduced mass, respectively. Here we have taken the $z$-axis for the direction of the incident wave and assumed that 
the scattering amplitude $f(\theta)$ depends only on the angle $\theta$. 

\begin{figure}[b]
\includegraphics[width=7.5cm]{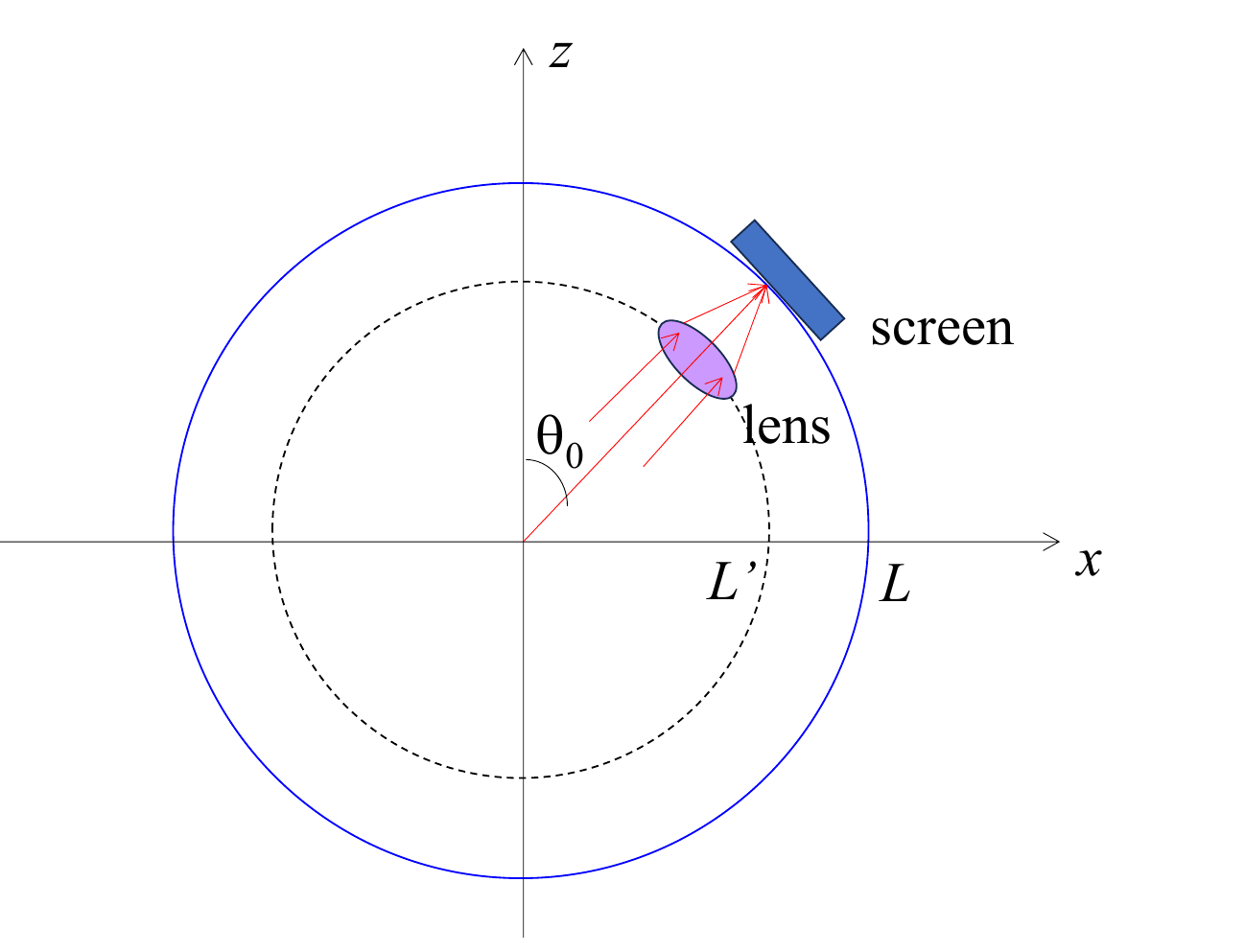}
\caption{
A schematic view of the set up of a lens and a screen for the imaging. The angle of the lens from the $z$ axis is $\theta_0$ and thus 
the distance of the lens from the $z$ axis is $L'\sin\theta_0$.
}
\label{fig:lens-screen1}
\end{figure}

\begin{figure}[b]
\includegraphics[width=8.5cm]{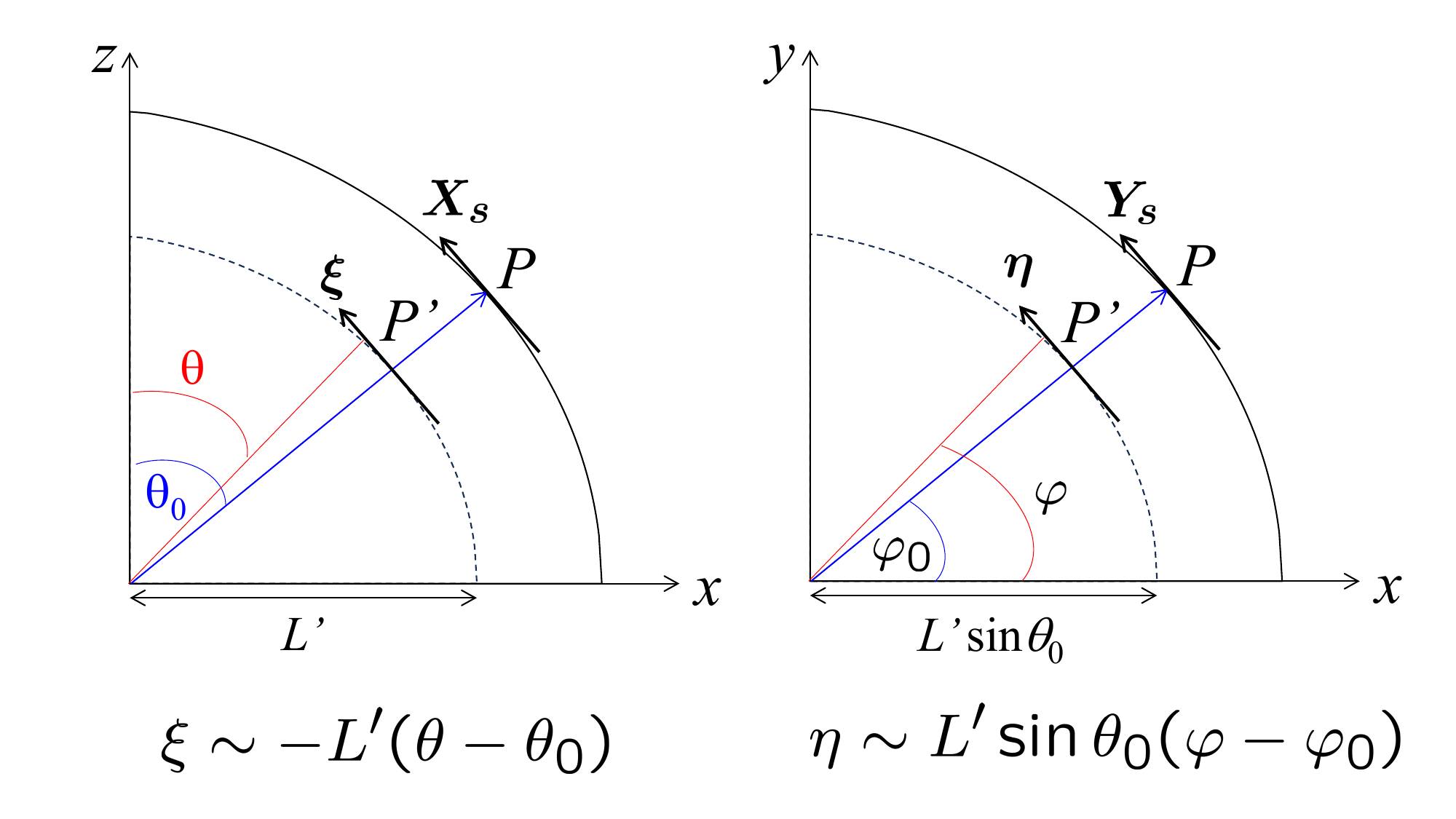}
\caption{
The definition of the coordinate systems $(\xi,\eta)$ and $(X_s,Y_s)$ for the imaging.  
The direction of $\xi$ and $X_s$ is taken to be in the $-\theta$-direction, while the direction of 
$\eta$ and $Y_s$ is in the $\varphi$-direction. 
}
\label{fig:lens-screen2}
\end{figure}

We put a convex lens at the distance $L'$ from the origin in the direction of $(\theta_0,\varphi_0)$ and take an image on the screen 
located at the distance $L$ from the origin (see Fig. \ref{fig:lens-screen1}). 
In Fig. \ref{fig:lens-screen2}, the center of the lens is denoted as $P'$, while the center of the screen is denoted as $P$, both of them 
are in the direction 
$(\theta_0,\varphi_0)$ from the origin. We use the two-dimensional 
Cartesian coordinate systems $(\xi,\eta)$ and $(X_s,Y_s)$ to express the position of a point on the lens and the screen, respectively. 
We put the lens in the tangential direction of the sphere at the point $P'$ and take the $\xi$ and the $\eta$ axis in the 
 $-\theta$ and the $\varphi$ directions, respectively. $\xi$ and $\eta$ are then expressed as 
$\xi\sim -L'(\theta-\theta_0)$ and $\eta\sim (L'\sin\theta_0)(\varphi-\varphi_0)$, respectively, for large values of $L'$. 

We assume that $L'$ is much larger than the size of the lens such that the wave which is 
incident on the lens can be approximately 
regarded as a plane wave. The role of the lens is to convert a plane wave to an incoming spherical wave (see Fig. 4 in Ref. \cite{hashimoto2020}). 
Assuming that the lens is infinitely thin, the amplitude at the point $(X_s,Y_s)$ on the screen then reads, 
\begin{eqnarray}
  \Psi_s(X_s,Y_s)&=&\int^{d_\xi}_{-d_\xi}d\xi\int^{d_\eta}_{-d_\eta}d\eta\,A(\xi,\eta) \nonumber \\
  &&\times e^{-ik[(X_s-\xi)^2+(Y_s-\eta)^2+(L-L')^2]^{1/2}},
\label{eq:strength}
\end{eqnarray}
where $A(\xi,\eta)$ is the amplitude for the scattering wave at the point $(\xi,\eta)$ on the lens, and 
the size of the lens is taken to be $d_\xi\times d_\eta$. 
We further assume that the size of the lens is much smaller than $L-L'$.  Eq. (\ref{eq:strength}) is then transformed to 
\begin{eqnarray}
  \Psi_s(X_s,Y_s)&\sim&
e^{-ik(L-L')}e^{-ik\frac{X_s^2+Y_s^2}{2(L-L')}} \nonumber \\ 
&&\times  
  \int^{d_\xi}_{-d_\xi}d\xi\int^{d_\eta}_{-d_\eta}d\eta\, 
  e^{ik\frac{\xi X_s+\eta Y_s}{L-L'}}
A(\xi,\eta).
\end{eqnarray}
Using the relations $\xi\sim -L'(\theta-\theta_0)$ and $\eta\sim L'\sin\theta_0(\varphi-\varphi_0)$, and by subsituting 
the scattering amplitude $f(\theta)$ to $A(\xi,\eta)$, one finds 
\begin{eqnarray}
  \Psi_s(X_s,Y_s)&\sim&
e^{-ik(L-L')}e^{-ik\frac{X_s^2+Y_s^2}{2(L-L')}} \nonumber \\ 
&&\times  
  \int^{\theta_0+\Delta\theta}_{\theta_0-\Delta\theta}d\theta\int^{\varphi_0+\Delta\varphi}_{\varphi_0-\Delta\varphi}d\varphi\,  
  \nonumber \\
&&  \times e^{ik\frac{-L'(\theta-\theta_0)X_s+L'\sin\theta_0Y_s(\varphi-\varphi_0)}{L-L'}}
f(\theta), 
\end{eqnarray}
with $\Delta\theta=d_\xi/L'$ and $\Delta\varphi=d_\eta/(L'\sin\theta_0)$. 
Introducing scaled coordinates $X\equiv -L'X_s/(L-L')$ and $Y\equiv L'\sin\theta_0 Y_s/(L-L')$, 
one finally obtains Eq. (1), up to a phase functor. 
Notice that the relation $\Delta\varphi=\Delta\theta/\sin\theta_0$ holds for a square lens, $d_\xi=d_\eta$.

\subsection{The image of Rutherford scattering}

We evaluate the image in the $X$ direction, 
\begin{equation}
   \Phi(X)=\int^{\theta_0+\Delta\theta}_{\theta_0-\Delta\theta}
   d\theta\,e^{ik(\theta-\theta_0)X}f(\theta), 
    \label{eq:image-supp}
\end{equation}
for small values of $\Delta\theta$. To this end, we expand $e^{ik(\theta-\theta_0)X}$ and $f(\theta)$ 
around $\theta=\theta_0$ up to the second order:
\begin{eqnarray}
    e^{ik(\theta-\theta_0)X}&\sim& 1+ik(\theta-\theta_0)X-k^2X^2(\theta-\theta_0)^2/2, \\
    f(\theta)&\sim&f(\theta_0)+f'(\theta_0)(\theta-\theta_0)+f''(\theta_0)(\theta-\theta_0)^2/2. \nonumber \\
\end{eqnarray}
The integral in Eq. (\ref{eq:image-supp}) can then be performed easily and reads,
\begin{eqnarray}
   \Phi(X)
   &\sim&2\Delta\theta\left\{f(\theta_0) \right.\nonumber \\ 
   &&\left.+\frac{(\Delta\theta)^2}{3}
\left(-\frac{k^2X^2}{2}f(\theta_0)+ik Xf'(\theta_0)+\frac{f''(\theta_0)}{2}\right)\right\}.    \nonumber \\
\end{eqnarray}
From this equation, one obtains 
\begin{eqnarray}
    \frac{d}{dX}|\Phi(X)|^2&\propto& -2k^2|f(\theta_0)|^2X \nonumber \\
    &&+ik(f^*(\theta_0)f'(\theta_0)-f(\theta_0)f'(\theta_0)^*). \nonumber \\
\end{eqnarray}
The peak of the image then appears at 
\begin{equation}
X=\frac{i}{2k}\left(\frac{f'(\theta_0)}{f(\theta_0)}-\frac{f'(\theta_0)^*}{f^*(\theta_0)}\right).
\end{equation}
We apply this to Rutherford scattering, whose scattering amplitude is given by, 
\begin{equation}
f_C(\theta)=-\frac{\eta}{2k\sin^2\frac{\theta}{2}}\,\exp\left[-i\eta\ln\left(\sin^2\frac{\theta}{2}\right)+2i\sigma_0\right]. 
\label{eq:coul}
\end{equation}
Here, $\eta=Z_1Z_2e^2/\hbar v$ is the Sommerfeld parameter, where $v$ is the relative velocity, and $\sigma_0={\rm arg}\Gamma(1+i\eta)$ 
is the $s$-wave Coulomb phase shift. 
For Eq. (\ref{eq:coul}), one finds 
\begin{equation}
\frac{f_C'(\theta)}{f_C(\theta)}=-(1+i\eta)\cot\left(\frac{\theta}{2}\right).
\end{equation}
The peak of the image therefore appears at 
\begin{equation}
    X=\frac{\eta}{k}\cot\left(\frac{\theta_0}{2}\right), 
\end{equation}
that is nothing but the impact parameter for Rutherford scattering. 

\subsection{The image of Mott scattering}

\begin{figure}[t]
\includegraphics[width=7.5cm]{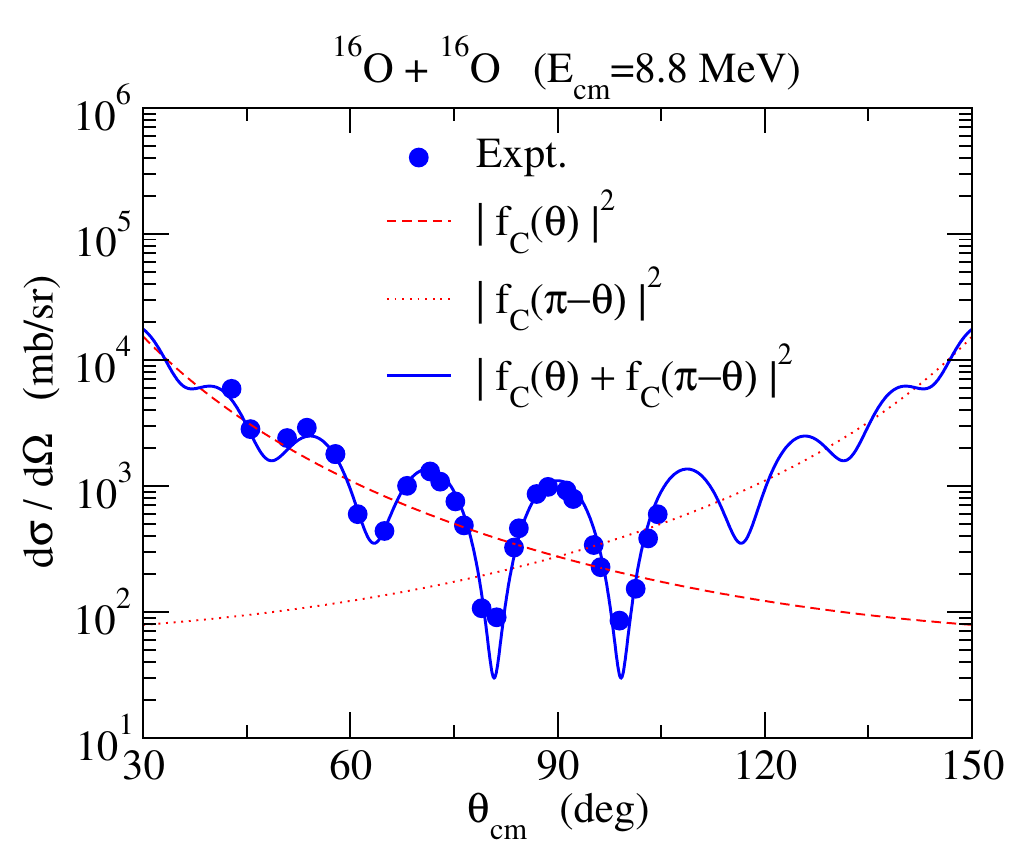}
\includegraphics[width=7.5cm]{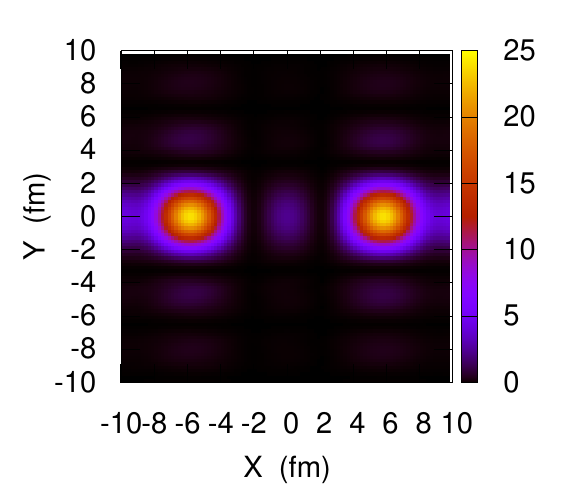}
\caption{
(Upper panel) The differential cross sections for elastic scattering of 
$^{16}$O+$^{16}$O at $E_{\rm c.m.}=8.8$ MeV. 
The contributions of unsymmetrized scattering amplitudes are also shown by the 
dashed and the dotted lines. The experimental data are taken from 
Ref. \cite{bromley1961}. 
(Lower panel) The image of Mott scattering shown in the upper panel. 
$\theta_0$ and $\Delta\theta$ are taken to be 90 and 30 degrees, respectively. 
}
\end{figure}

Let us consider Mott scattering, i.e., scattering of two identical particles, for which 
the scattering amplitude is given by $f(\theta)\pm f(\pi-\theta)$. 
For the component $f(\pi-\theta)$, the $X$ dependence of Eq. (1) reads
\begin{eqnarray}
\Phi_{\pi-\theta}(X)&\equiv&\int^{\theta_0+\Delta\theta}_{\theta_0-\Delta\theta}
d\theta\, e^{ik(\theta-\theta_0)X} f(\pi-\theta) \\
&=& 
\int^{\pi-\theta_0+\Delta\theta}_{\pi-\theta_0-\Delta\theta}
d\tilde{\theta}\, e^{ik(\pi-\tilde{\theta}-\theta_0)X} f(\tilde{\theta}),
\end{eqnarray}
where $\tilde{\theta}$ is defined as $\tilde{\theta}=\pi-\theta$. 
For $\theta_0=\pi/2$, this is equivalent to 
\begin{eqnarray}
\Phi_{\pi-\theta}(X)
&=& 
\int^{\theta_0+\Delta\theta}_{\theta_0-\Delta\theta}
d\tilde{\theta}\, e^{ik(\theta_0-\tilde{\theta})X} f(\tilde{\theta})=\Phi_{\theta}(-X). \nonumber \\
\end{eqnarray}
Thus, the image of Mott scattering is symmetric with respect to $X=0$, and it therefore has 
two symmetric peaks at $X_{\rm peak}$ and $-X_{\rm peak}$. 

The upper panel of Fig. 7 shows the differential cross sections for elastic scattering of 
$^{16}$O+$^{16}$O at $E_{\rm c.m.}=8.8$ MeV. 
This energy is at about 1.5 MeV below the Coulomb barrier, and the nuclear effect can be 
neglected. In fact, the experimental data can be well fitted using the Coulomb scattering amplitudes, 
$d\sigma/d\Omega=|f_C(\theta)+f_C(\pi-\theta)|^2$. 
The contributions of $f_C(\theta)$ and $f_C(\pi-\theta)$ are also shown by the dashed and the dotted 
lines, respectively. The image of Mott scattering is shown in the lower panel. 
$\theta_0$ and $\Delta\theta$ are taken to be 90 and 30 degrees, respectively. 
As we have argued, 
the image has two symmetric peaks. A comparison with Fig. 1 indicates that the peak at a positive $X$ 
corresponds to the contribution of $f_C(\theta)$, while the peak at a negative $X$ 
corresponds to the contribution of $f_C(\pi-\theta)$. 

\bibliography{ref}
\bibliographystyle{apsrev4-2}